\documentclass[12pt]{article}

\begin{document}
\input {epsf}

\newcommand{\beq}{\begin{equation}}
\newcommand{\eeq}{\end{equation}}
\newcommand{\beqa}{\begin{eqnarray}}
\newcommand{\eeqa}{\end{eqnarray}}

\def\ov{\overline}
\def\onlyif{\rightarrow}

\def\openone{\leavevmode\hbox{\small1\kern-3.8pt\normalsize1}}

\def\a{\alpha}
\def\b{\beta}
\def\g{\gamma}
\def\r{\rho}
\def\minus{\,-\,}
\def\eks{\bf x}
\def\kay{\bf k}

\def\ket#1{|\,#1\,\rangle}
\def\bra#1{\langle\, #1\,|}
\def\braket#1#2{\langle\, #1\,|\,#2\,\rangle}
\def\proj#1#2{\ket{#1}\bra{#2}}
\def\expect#1{\langle\, #1\, \rangle}
\def\trialexpect#1{\expect#1_{\rm trial}}
\def\ensemblexpect#1{\expect#1_{\rm ensemble}}
\def\kpsi{\ket{\psi}}
\def\kphi{\ket{\phi}}
\def\bpsi{\bra{\psi}}
\def\bphi{\bra{\phi}}

\def\ditto{\rule[0.5ex]{2cm}{.4pt}\enspace}
\def\th{\thinspace}
\def\ni{\noindent}
\def\thirty{\hbox to \hsize{\hfill\rule[5pt]{2.5cm}{0.5pt}\hfill}}

\def\set#1{\{ #1\}}
\def\setbuilder#1#2{\{ #1:\; #2\}}
\def\Prob#1{{\rm Prob}(#1)}
\def\pair#1#2{\langle #1,#2\rangle}
\def\Id{\bf 1}

\def\dee#1#2{\frac{\partial #1}{\partial #2}}
\def\deetwo#1#2{\frac{\partial\,^2 #1}{\partial #2^2}}
\def\deethree#1#2{\frac{\partial\,^3 #1}{\partial #2^3}}

\def\openone{\leavevmode\hbox{\small1\kern-3.8pt\normalsize1}}

\title{Quantum Seals}
\author{
H. Bechmann-Pasquinucci\thanks{Permanent address: {\rm UCCI.IT}, {\it via
Olmo 26,
I-23888 Rovagnate (LC), Italy.}\hspace{3cm}\break{\rm email: 
bechmann@ucci.it}}
\\
\small
{\it University of Pavia, Dipartimento di Fisica 'A. Volta', Via Bassi 6, 
27100 Pavia, Italy}}
\date{June 25, 2003}
\maketitle

\abstract{
A quantum seal is a way of 
encoding 
a classical message into quantum states, so that everybody can read 
the message error-free, but at the same time the sender and all intended 
readers who have some prior knowledge of the quantum seal, can check if 
the 
seal has been broken and the message read. The verification is done 
without reading  nor disturbing the sealed 
message. 
} 
\normalsize

\section {Introduction} 
Before the age of electronic transfer of information, important 
letters and documents were often closed using a wafer of molten wax into 
which was pressed the distinctive seal of the sender \cite{cs}. This was 
meant to 
fulfill different purposes, namely the authentication of the sender, but it 
also enabled the receiver to verify that the letter or document had not 
been opened and read by a third party. This is refered to as a classical 
seal.

It is important to notice the following points. The sender is not 
committed to the content, since she can always change her mind, 
write a new letter and seal it. Moreover the receiver needs to 
have  
prior knowledge of the symbol of the seal in order to verify that it is 
actually authentic --- but still it could be falsified and in principle 
the only person who can truly verify that the seal and the content is 
authentic is the sender. Even further, the seal can be broken by anyone 
who 
wishes to learn the content of the letter, by simply physically breaking 
the 
seal, and then open and read the letter.

In this paper the  idea for making a Quantum Seal is presented. The 
protocol 
belongs to the same category as quantum cryptography \cite{qc}, quantum 
bit 
commitment \cite{qbc1, qbc2}, quantum signature \cite{qs} and 
authentication of 
quantum messages \cite{aqm} etc. For the quantum seal  
a classical message is encoded into quantum states, in such a way 
that the quantum seal can be broken by anyone and the message read 
error-free. However  it is possible to verify  if the message has 
been 
read 
by checking if the quantum seal 
has been broken. The first goal is that this verification can be 
performed by the sender. Ultimately 
the quantum seal will possess the same properties as the classical seal, 
which means that anyone with some prior knowledge about the quantum seal 
will be able to check if it has been broken. 

A first implementation of the protocol for quantum seals will be 
presented.
The encoding of  the message is performed by using the simplest form 
of 
classical error correction codes \cite{cec}. And the verification of the 
quantum seal is
performed by using the so-called 'SWAP-test' \cite{st}.
Eventhough this implementation is only resistant to single qubit attacks 
(see section 5) 
it serves as a 
good 
illustration of the  properties of quantum seals.

\section{Encoding and verification by Alice}
The first step in order to create the quantum seal is to find a way of 
encoding a classical message into quantum states so that the 
encoder, Alice, can always verify if the message has been read, i.e. she 
can check if the quantum seal has been broken. At the same time anyone 
can actually read the message. This can be achieved in a very simple way. 

A classical message can be encoded into quantum states and read perfectly 
without errors by anyone if Alice writes the message using qubits 
states always prepared in the same basis and then announces which basis 
she 
used. For example, if she wants to write a string of bits, then she could 
use the states $\ket{0}\def \ket{0_z}$ and $\ket{1}\def\ket{1_z}$ 
corresponding to the 
$z$ (computational) 
basis to signify 0 and 1 respectively, and then announce the basis she  
used. This will allow anyone to measure the qubits in the correct basis 
and read the message. But in this way, naturally Alice have no way of 
checking if someone actually {\it did} read the message. 

In order for Alice to be able to check if the message has been read, it is
necessary to add states from a different basis, which if measured in the
wrong basis will lead to errors --- similarly to what is done in 
quantum cryptography. Still, since everybody should be able to 
read the message this has to be done in a clever way. 

This can be done by encoding each classical bit into a 
block of three qubits in such a way that when measured in the basis 
announced by Alice, the encoding is self-correcting, which means that the 
bit can be learned without error. However, at the same time Alice can 
check that it has been 
read. This can be achieved in the following way: Two of the qubits will be 
prepared in the same state in the computational basis according to the bit 
value Alice  wants to write, the third qubit is a control qubit which is 
prepared in a state from one of the mutually unbiased bases $x$ and $y$.
In other words two of the qubits are prepared in the message reading-basis 
($\ket{0}$ and $\ket{1}$), whereas one control qubit will be in one 
of the four states
$\ket{0_x}$, $\ket{1_x}$, $\ket{0_y}$ or $\ket{1_y}$. The state 
and position of the 
control qubit is chosen at random by Alice --- the choice is done 
independently for each block of qubits. This means that in each block of 
three qubits two qubits are in the correct state, for example if Alice 
wants to write bit value 0, two of the qubits will be 
in the state $\ket{0}$, whereas the state of the last qubit will be chosen 
at random 
between the 
states from the other mutually unbiased bases.

Now suppose Alice wants to write the classical bit sequence $0110 ...$, 
then there are many ways for her to encode it, one way would be the 
following:
\begin{eqnarray}
&&\ket{0}\ket{0}\ket{0_y}~~~~~{\rm bit~number~1,~value}=0 \nonumber\\
&&\ket{0_x}\ket{1}\ket{1}~~~~~{\rm bit~number~2,~value}=1 \nonumber\\
&&\ket{1}\ket{0_y}\ket{1} ~~~~~{\rm bit~number~3,~value}=1 \nonumber\\
&&\ket{0}\ket{0_x}\ket{0} ~~~~~{\rm bit~number~4,~value}=0 \nonumber\\
&&... \nonumber
\end{eqnarray}
where each triplet of states corresponds to the encoding of one bit. 
Since Alice has announced in public which basis is the message 
reading-basis 
(here chosen to be the computational basis, i.e. $z$ basis), 
anyone can read it, but it will automatically introduce errors which can 
be identified by Alice. But since 
the encoding is self-correcting the message can nevertheless be read 
without errors. 

To see this, assume that the reader is always measuring in the correct 
message reading-basis, 
then naturally all the states which are encoded in this 
basis will be read perfectly without error. But in each triplet there is 
one state which belongs to a basis which is mutually unbiased with the 
reading-basis, and for this state the bit value 0 or 1 is found with equal 
probability. For example $\ket{0}\ket{0}\ket{0_y}$, when read in the 
computational basis, could result in either $000$ or $001$ --- with equal 
probability. However, since in each triplet there is only one control 
qubit, there can be at maximum one error and hence the correct bit value 
can be obtained by a simple majority vote. Which means that the message 
can be read without error. 

On the other hand since the control qubit now 
has been measured in the reading-basis, a subsequent measurement in the 
correct preparation basis (in this case the $y$-basis) will
lead to an error, with  
probability $1/2$. This will allow Alice at any stage to 
check 
if the message has been read, because she can at any time measure each 
qubit in its preparation basis and if she finds errors she will conclude 
that the seal has been broken and the message read.

Notice an interesting thing, namely that it is only with probability $1/2$ 
that the reader will learn the position of the control qubit.  
In 
the example described above, with probability $1/2$, the reader will find 
$000$ from which the position of the control bit can not be identified, 
whereas with probability $1/2$ the result will be  $001$, from which it is 
obvious that the last bit corresponds to the position of the control qubit 
--- but, of course, the reader learns nothing of the original state of the 
control qubit.

This part of the protocol enables everyone to read the message written by 
Alice error free, but at the same time it enables Alice to check if the 
quantum seal 
has been broken, because the message can not be read without giving rise 
to errors.

\section{Intended reader verification}
At this point in the protocol, Alice has written and sealed (encoded) her
message into triplets of qubit states and stored them in a quantum memory
which is accessible by anyone. Which means that the sealed message is now 
located in a public area. Notice that the sealed message is just a big 
product 
state of $N$ qubits, $\ket{{\phi}_{1}}\ket{{\phi}_{2}}\ket{{\phi}_{2}}
\cdots\ket{{\phi}_{N}}$, where $\ket{{\phi}_{i}}$ is the state of the 
$i'th$ qubit and where $N=3\times({\rm number~of ~bits})$.

In order to have complete correspondence with the classical seal, not only 
Alice should be able to verify that the seal has not been broken, but all 
intended readers should be able to verify that the quantum seal is still 
intact, without reading, nor disturbing the message written by Alice.

In the case of the classical seal the receiver is familiar with the design 
of the symbol pressed into the wax, and uses this knowledge to identify 
the 
letter received as authentic. This means that prior knowledge is required 
in the classical case. For the quantum seal, this part of the protocol is 
obtained by Alice preparing additional qubits which she then distributes 
to 
the various intended readers, who can then use them to verify if the 
quantum seal is still intact. 

The additional qubits prepared by Alice correspond to a subset of the 
message and control qubits. Alice does not reveal the state of any of the 
qubits, but merely informs the receiver which qubits they correspond to in 
the sealed message. For example Bob-1, 
gets 
copies\footnote{Alice can, of course, produce  any 
number of copies of a given state, since she knows the state of any given 
qubit. So this does not go against the non-cloning theorem.} 
 of qubit state number 1, 7, 16, 30, etc. but he doesn't know the 
state, nor does he know if a given qubit is a message qubit or a control 
qubit (this issue will be addressed further below). Similarly, all the 
other 
intended readers (Bob-2 to 
Bob-L) likewise receive a set of qubits corresponding to a subset of the 
message and control qubits.

The verification of the quantum seal as such is done by performing a 
so-called SWAP test. A SWAP test can be used to evaluate if two states are 
identical, without any knowledge of the states themselves. This means that 
Bob-1, for example, can use his subset to make a SWAP test with the 
corresponding qubits in the sealed message. 
If he finds complete agreement, i.e. if all his qubits pass the SWAP test, 
he will conclude that the quantum seal is still intact and the message 
has not been read. Whereas if he finds that just one of his qubits fail 
the SWAP test he will conclude that the seal has been broken and the 
message has been read. 

The SWAP test works in the following way: Bob chooses the qubit he wants 
to 
test, $\ket{{\phi}_{test}}$, and takes the corresponding qubit 
from the sealed message, $\ket{{\phi}_{seal}}$, (remember Bob knows which 
number it corresponds to, for example, qubit number 1). Then he couples 
the 
two qubits with an additional qubit, an ancilla, in the state $\ket{0}$, 
i.e. Bob now has the three qubit quantum state 
$\ket{0}\ket{{\phi}_{test}}\ket{{\phi}_{seal}}$, and he performs the 
following operations  one the full system
\begin{eqnarray}
&&\left({\rm H}\otimes\openone\otimes\openone\right)\left({\rm 
c-SWAP}\right)
\left({\rm H}\otimes\openone\otimes\openone\right)~~~
\ket{0}\ket{{\phi}_{test}}\ket{{\phi}_{seal}}\nonumber \\
&&= {\frac{1}{2}}\ket{0}\left(\ket{{\phi}_{test}}\ket{{\phi}_{seal}}+ 
\ket{{\phi}_{seal}}\ket{{\phi}_{test}}\right)\nonumber \\
&&+{\frac{1}{2}}\ket{1}\left(\ket{{\phi}_{test}}\ket{{\phi}_{seal}}-
\ket{{\phi}_{seal}}\ket{{\phi}_{test}}\right)
\end{eqnarray}
where H is the Hadamard transformation, $\openone$ the identity operator 
and SWAP is the transformation: 
$\ket{{\phi}_{test}}\ket{{\phi}_{seal}} \rightarrow
\ket{{\phi}_{seal}}\ket{{\phi}_{test}}$ and c-SWAP means that it is a 
SWAP which is controlled by the ancilla (for a more detailed description 
of the SWAP test see \cite{st}). These transformations are 
followed 
by a measurement of the ancilla in the $\ket{0}$, $\ket{1}$ basis. The 
state $\ket{1}$ is found with probability 
${\frac{1}{2}}-{\frac{1}{2}}{|\braket{{\phi}_{test}}{{\phi}_{seal}}|^2}$, 
from which it is seen that if $\ket{{\phi}_{test}}=\ket{{\phi}_{seal}}$, 
the ancilla will never be found in the state $\ket{1}$, but $\ket{0}$ will 
always be obtained. In other words if Bob obtains $\ket{1}$, he will conclude 
that the states were not identical, hence they did not pass the SWAP test.

If the sealed message has not been read, all of Bob's qubits will pass the 
SWAP 
test, whereas if the message has been read some of Bob's qubit will fail 
the SWAP test and he will consequently conclude that the quantum seal has 
been broken and assume that the message has been read.

Suppose that Bob's test qubit is in one of the four states from the 
$x$ 
or the $y$ basis, which means that it is one of the control qubits; and 
assume  that 
the message has been read in the $z$ basis (as announced by Alice) then 
the control qubit is no longer in its correct state, but  in one of the 
states from the $z$ basis. If Bob performs a SWAP test, then, since 
all the involved states are mutually unbiased, the probability that the 
SWAP test will fail is 
${\frac{1}{2}}-{\frac{1}{2}}{|\braket{{\phi}_{test}}{{\phi}_{seal}}|^2}=
{\frac{1}{2}}-{\frac{1}{2}}{\frac{1}{2}}={\frac{1}{4}}$. This means that 
if Alice supplies Bob with sufficiently many qubits, so that he can  
perform many SWAP tests, statistically Bob should obtain an error and 
conclude that the seal has been broken.

Notice one important point namely that the SWAP test does not give Bob any 
information about the state, nor will it destroy the seal if it is intact. 
Which exactly satisfies the intended reader verification requirements: the 
quantum seal can 
be verified without reading nor disturbing the sealed message.

\section{Some security aspects}
A fully detailed security analysis of the protocol for quantum seals is 
beyond the scope of the present paper. Indeed there may exist many 
different implementations of the idea and, most probably, each 
implementation must be analyzed independently. Here I mention a few security 
aspects 
related to the presented implementation of the protocol, which may also 
prove 
important to future implementations, and in the next section discuss the 
limitation of the proposed implementation.

{\bf The encoding:} Someone may attempt to read the sealed message by 
reading only two out of three qubits in a triplet, hoping to obtain the 
same result twice and hence know the bit value without reading all three 
qubit states. The hope would be to avoid reading the control 
qubit 
and hence avoid making any errors. But if the sealed message is 
sufficiently 
long, errors should be  guaranteed by statistics.

{\bf The distribution of subsets of qubits:} 
If Alice provides the 
intended 
readers with the number 
of the qubits in public this could be exploited by others to avoid reading 
exactly those qubits. This means that each reader should know only his 
subset of qubits and not the subset of any other. However, this problem 
could easily be solved by Alice sending this information to each of the 
intended readers using quantum cryptography.

{\bf The nature of the subsets:} In each subset
some of the qubits should correspond to message qubits and some will
correspond to control qubits. In principle the message qubits can not be 
used to check the quantum seal (they will always pass the SWAP test) only 
the control qubits can be used for 
that. However each intended reader receives both message qubits and 
control 
qubits to avoid that they can use the knowledge of the position of the 
control qubits to cheat. 
The number of qubits supplied to each reader, 
should be such that by simply reading the subset he has received  he can 
not not learn the 
message, but at the same time he must possess enough qubits in order to 
actually 
check if the quantum seal is still intact (in terms of good statistics). 

\section{Limitations of the proposed implementation}
Above it was shown how someone trying to read the sealed message by 
measuring each qubit independently will automatically introduce errors, 
hence break the quantum seal and unavoidedly be detected. 

However, for the proposed implementation, it is possible to completely 
avoid detection by making collective
measurement on each triplet of qubits, hence reading the sealed message 
bit
by bit (remember that each bit is encoded into three qubits). This is due 
to the fact that the twelve different three qubit states which encode bit 
value 0 are orthogonal to  the twelve three qubit states which encode bit 
value 1. Indeed all the 0-states lie in the subspace spanned by the 
following states (in the z-basis), 
$\ket{000}$, $\ket{001}$, $\ket{010}$ and $\ket{100}$, whereas the 
1-states 
all lie in the orthogonal subspace spanned by  
$\ket{111}$, $\ket{110}$, $\ket{101}$ and $\ket{011}$. This means that a 
measurement of the corresponding projectors:
\begin{eqnarray}
P_0=\proj{000}{000}+\proj{001}{001}+\proj{010}{010}+\proj{100}{100}\nonumber\\
P_1=\proj{111}{111}+\proj{110}{110}+\proj{101}{101}+\proj{011}{011}
\end{eqnarray}
will distingush perfectly between the 0 and 1-states without disturbing 
the states. Which means that this kind of collective measurement will 
allow someone 
to read the sealed message without introducing any errors and hence 
will avoid detection.

In other words  the presented implementation of the protocol is not 
robust
against collective measurements, but only against single qubit attacks.
However the described implementation serves as a perfect example and
illustration of the proporties of the protocol for quantum seals.

\section{Concluding remarks}
Quantum seals have been introduced, they can be used to verify 
if a classical message encoded into quantum states have been read. A first 
implementation of this idea in terms of classical 
error correcting codes and a so-called SWAP test has been presented. 
It has been shown that someone trying to read the sealed message by 
measuring each qubit independently will automatically introduce errors and 
hence break the quantum seal and be detected. 

However, the proposed 
implementation is resistant only against single qubit attacks. Someone who 
is able to make a collective measurement on three qubits will be able to 
read the sealed message without 
making any errors. This means that it is an open problem to 
find an implementation of the protocol for quantum seals which offers 
protection against any 
kind of attack. A completely secure implementation of quantum seals may 
very well require different tools than the classical error correcting 
codes and the SWAP test which is used here, for example one could imagine 
the use of entanglement and Bell Inequalities.

It should be emphasized that Alice is not committed to the content of her
message, as for the classical seal she is absolutely free to change her
mind, write a new message and seal it.  Moreover, one of the basic ideas
is that anyone should be able to read the message written by Alice, which
means that it is not the security of the content which has to be accounted
for, but instead the guarantee that anyone reading the message will
actually leave a trace, i.e. some errors.

Finally notice the following interesting points; The protocol for quantum
seals require no classical communication between Alice and the intended
readers contrary to most other protocols in the area of quantum
cryptography. A priory it only requires that Alice is able to send quantum
states and make public (classical) announcements.  More, the classical
seal is a physical object namely a wax wafer, whereas the quantum seal is
somehow a way of encoding. I do not know of any protocol which can be
implemented, for example, on a classical computer which works as a seal.

\section*{Acknowledgements}
I would like to thank Andrea Pasquinucci and Adrian Kent for many 
stimulating discussions. This work has been supported by the FET European 
Networks on Quantum Information and Communication Contract 
IST-2000-29681:ASESIT.


\begin{thebibliography}{99}
\bibitem{cs} Can be bought in departments stores and paper shops, to 
be used for old fashion letter writing or gift wrapping.
\end{thebibliography}
\end{document}